\documentclass[11pt]{article}

\usepackage{amsmath}
\usepackage{amssymb}
\usepackage{amsthm}
\usepackage{multirow}
\usepackage{color}
\usepackage{longtable}
\usepackage{array}
\usepackage{url}
\usepackage{tikz}

\newtheorem{theorem}{Theorem}[section]
\newtheorem{definition}{Definition}[section]
\newtheorem{lemma}[theorem]{Lemma}

\title{New constructions of quantum MDS convolutional codes derived from generalized Reed-Solomon codes
\footnote{The research of G. Ge was supported by the National Natural Science Foundation of China under Grant Nos. 61171198, 11431003 and 61571310, and the Importation and Development of High-Caliber Talents Project of Beijing Municipal Institutions.}
}

\author{
Baokun Ding \thanks{B. Ding is with the School of Mathematical Sciences, Zhejiang University,
Hangzhou 310027,  China (e-mail: bkding@zju.edu.cn).}
\and Tao Zhang \thanks{T. Zhang is with the School of Mathematical Sciences, Zhejiang University,
Hangzhou 310027,  China (e-mail: tzh@zju.edu.cn).}
\and Gennian Ge
\thanks{G. Ge is with  the School of Mathematical Sciences, Capital Normal University,
Beijing 100048, China. He is also with Beijing Center for Mathematics and Information Interdisciplinary Sciences, Beijing 100048, China (e-mail: gnge@zju.edu.cn).}
}

\begin{document}

\date{}\maketitle

\begin{abstract}
Quantum convolutional codes can be used to protect a sequence of qubits of arbitrary length against decoherence. In this paper, we give two new constructions of quantum MDS convolutional codes derived from generalized Reed-Solomon codes and obtain eighteen new classes of quantum MDS convolutional codes.
Most of them are new in the sense that the parameters of the codes are different from all the previously known ones.
\medskip

\smallskip
\end{abstract}

\section{Introduction}
In recent years, there has been tremendous interest in constructing quantum block codes, which is an important subject in quantum information and quantum computing. Quantum convolutional codes give quantum block codes an alternative to protect quantum information for reliable quantum communication over noisy quantum channels. And quantum convolutional codes can also be used to protect a sequence of qubits of arbitrary length against decoherence. Therefore, constructing good quantum convolutional codes has been an important research problem.

The original definition of quantum convolutional codes appeared in \cite{def1}. Afterwards, Ollivier and Tillich \cite{Harold2003Description,Ollivier2004Quantum} presented the fundamentals of quantum convolutional codes and explained how to encode a stream of qubits efficiently. Quantum convolutional codes obtained from classical convolutional codes are provided in \cite{Forney2007Convolutional,Lidar2013Quantum}. Tan and Li \cite{Tan2010Efficient} proposed five systematic constructions by using LDPC and LDPC-convolutional codes. La Guardia constructed several new families of quantum negacyclic MDS convolutional codes in \cite{Guardia2013On} and unit-memory quantum convolutional BCH codes in \cite{Giuliano2014On}. Some new nonbinary quantum convolution codes are also obtained in \cite{CLHL,ZCL} by using constacyclic codes. Further results about quantum convolutional codes can be found in \cite{De2004A,Grassl2006Non,Grassl2007Constructions,Grassl2007Quantum,QJZ,Wilde2007Entanglement,Wilde2010Convolutional}.

Generalized Reed-Solomon codes (GRS codes) are an important group of error correcting codes. They are MDS (maximum distance separable) codes as their parameters meet the Singleton bound. Recently, they were applied to construct quantum codes. In \cite{JLLX}, Jin et al.\ used GRS codes and algebraic geometry codes to construct quantum MDS codes. In \cite{Aly2007Quantum}, Aly et al. presented some quantum MDS convolutional codes from GRS codes based on limit constructions. Some pure asymmetric quantum MDS codes from GRS codes were given in \cite{EJKL}. By using GRS codes, Zhang et al.\ \cite{ZG15} obtained some new constructions of $q$-ary quantum MDS codes. It seems that GRS codes are a rich source of constructing quantum MDS codes with good parameters.

In this paper, we present two new constructions of quantum MDS convolutional codes by using GRS codes. Consequently, we obtain eighteen new classes of quantum MDS convolutional codes. Most of them are new in the sense that the parameters of the codes are different from the ones available in the literature.

This paper is organized as follows. In Section \ref{secpre}, we recall some preliminary concepts and results about classical convolutional codes,  quantum convolutional codes and GRS codes. In Section \ref{seccon}, we propose new constructions of quantum MDS convolution codes derived from GRS codes. Section \ref{conclusion} concludes the paper.
\section{Preliminaries}\label{secpre}
In this section, we recall some basic notations and necessary facts which are important to our constructions. Throughout this paper, we always assume that $q$ is a prime power and $\mathbb{F}_{q}$ is the field with $q$ elements if not specified.

 A linear $[n,k]$ code $\mathcal{C}$ over $\mathbb{F}_{q}$ is a $k$-dimensional subspace of $\mathbb{F}_{q}^{n}$. The weight $\textup{wt}(x)$ of a codeword $x\in \mathcal{C}$ is the number of nonzero components of $x$. The distance of two codewords $x,y\in \mathcal{C}$ is $d(x,y)=\textup{wt}(x-y)$. The minimum distance $d$ of $\mathcal{C}$ is the minimum distance between any two distinct codewords of $\mathcal{C}$. An $[n,k,d]_{q}$ code is an $[n,k]$ code over $\mathbb{F}_{q}$ with the minimum distance $d$.

Given two vectors $x=(x_{0},x_{1},\cdots,x_{n-1}),\ y=(y_{0},y_{1},\cdots,y_{n-1})\in\mathbb{F}_{q}^{n}$. We define the Euclidean inner product as $\langle x,y\rangle_{E}=\sum_{i=0}^{n-1}x_{i}y_{i}$. When $q=l^{2}$, where $l$ is a prime power, we also consider the Hermitian inner product which is defined by $\langle x,y\rangle_{H}=\sum_{i=0}^{n-1}x_{i}y_{i}^{l}$.
We define the Euclidean dual code of $\mathcal{C}$ as
$$\mathcal{C}^{\bot E}=\{x\in\mathbb{F}_{q}^{n}|\langle x,y\rangle_{E}=0\textup{ for all }y\in \mathcal{C}\}.$$
Similarly, the Hermitian dual code of $\mathcal{C}$ is defined as
$$\mathcal{C}^{\bot H}=\{x\in\mathbb{F}_{q}^{n}|\langle x,y\rangle_{H}=0\textup{ for all }y\in \mathcal{C}\}.$$
A linear code $\mathcal{C}$ is called Euclidean (Hermitian) dual-containg if $\mathcal{C}^{\bot E}\subseteq \mathcal{C}$ ($\mathcal{C}^{\bot H}\subseteq \mathcal{C}$, respectively). The following theorem gives a bound on the minimum distance of a linear code.

\begin{lemma}{\rm \cite{FUN}}
Every linear $[n,k,d]_{q}$ code $\mathcal{C}$ satisfies the Singleton bound
\begin{equation*}
d\leq n-k+1.
\end{equation*}
A code achieving this bound is called an MDS code.
\end{lemma}

\subsection{Classical Convolution Codes}
 Recall that a convolutional code $\mathcal{C}$ of length $n$ and dimension $k$ over $\mathbb{F}_{q^{2}}$ is a free module of rank $k$ that is a direct summand of $\mathbb{F}_{q^{2}}[D]^{n}$ and a polynomial encoder matrix $G(D)=(g_{ij})\in \mathbb{F}_{q^{2}}[D]^{k\times n}$ is called basic if it has a polynomial right inverse. A basic generator matrix is called reduced (or minimal \cite{FUN}) if the overall constraint length $\gamma=\Sigma_{i=1}^{k}\gamma_{i}$ has the smallest value among all basic generator matrices, where $\gamma_{i}=\max_{1\leq j \leq n}\lbrace \textup{deg}~g_{ij}\rbrace$; in this case the overall constraint length $\gamma$ will be called the degree of the code.

\begin{definition}{\rm \cite{Aly2007Quantum}}
A convolutional code $V$ with parameters $(n,k,\gamma; \mu,d_{f})_{q^{2}}$ is a submodule of $\mathbb{F}_{q^{2}}[D]^{n}$ generated by a reduced basic matrix $G(D)=(g_{ij})\in \mathbb{F}_{q^{2}}[D]^{k\times n}$, $V=\lbrace {\bf u}(D)G(D)|{\bf u}(D)\in \mathbb{F}_{q^{2}}[D]^{k}\rbrace$,where $n$ is the length, $k$ is the dimension, $\gamma=\Sigma_{i=1}^{k}\gamma_{i}$ is the degree, $\mu =\max_{1\leq i\leq k}\lbrace \gamma_{i}\rbrace$ is the memory and $d_{f}=wt(V)=\min\lbrace wt({\bf v}(D))|{\bf v}(D)\in V, {\bf v}(D)\neq0 \rbrace$ is the free distance of the code. Here, $wt({\bf v}(D))=\Sigma_{i=1}^{n}wt(v_{i}(D))$, where $wt(v_{i}(D))$ is the number of nonzero coefficients of $v_{i}(D)$.
\end{definition}

We define the Hermitian inner product on $\mathbb{F}_{q^{2}}[D]^{n}$ as $\langle{\bf u}(D),{\bf v}(D)\rangle_{H}=\Sigma_{i}{\bf u}_{i}\cdot{\bf v}_{i}^{q}$, where ${\bf u}_{i},{\bf v}_{i}\in \mathbb{F}_{q^{2}}^{n}$, ${\bf v}_{i}=(v_{1i},v_{2i},\cdots,v_{ni})$ and ${\bf v}_{i}^{q}=(v_{1i}^{q},v_{2i}^{q},\cdots,v_{ni}^{q})$. The Hermitian dual of the code V is defined by
\begin{equation*}
V^{\bot H}=\lbrace {\bf u}(D)\in \mathbb{F}_{q^{2}}[D]^{n}|\langle{\bf u}(D),{\bf v}(D)\rangle_{H}=0 \textup{ for all } {\bf v}(D)\in V\rbrace.
\end{equation*}

Let $\mathcal{C}$  be an $[n,k,d]_{q^{2}}$ linear code with parity check matrix $H$. Split $H$ into $\mu +1$ disjoint submatrices $H_{i}$ such that
\begin{equation}\label{eqH}
H=
\left(
  \begin{array}{c}
    H_{0}   \\
    H_{1} \\
    \vdots \\
    H_{\mu} \\
  \end{array}
\right),
\end{equation}
where each $H_{i}$ has $n$ columns. Then we obtain the polynomial matrix
\begin{equation}\label{GD}
G(D)=\tilde{H}_{0}+\tilde{H}_{1}D+\cdots +\tilde{H}_{\mu}D^{\mu},
\end{equation}
where the matrices $\tilde{H}_{i}$ are derived from the respective matrices $H_{i}$ by adding zero-rows at the bottom to ensure that the matrix $\tilde{H}_{i}$ has $\kappa$ rows for all $1 \leq i \leq \mu $, and here $\kappa$ denotes the maximal number of rows among the matrices $H_{i}$, for $1\leq i \leq \mu$. Then the matrix $G(D)$ with $\kappa$ rows generates a convolutional code and $\mu$ is the memory of the code.

\begin{theorem}{\rm \cite{Aly2007Quantum2}}\label{gen}
Suppose that $\mathcal{C}$ is a linear code over $\mathbb{F}_{q^2}$ with parameters $[n,k,d]_{q^{2}}$ and assume also that $H\in \mathbb{F}_{q^{2}}^{(n-k)\times n}$ is a parity check matrix for $\mathcal{C}$ partitioned into submatrices $H_{0},H_{1},\cdots,H_{\mu}$ as in (\ref{eqH}) such that $\kappa=rank({H_{0}})$ and $rank(H_{i})\leq \kappa$ for $1\leq i\leq \mu$ and consider the polynomial matrix $G(D)$ as in (\ref{GD}). Then we have:
\\(a) The matrix $G(D)$ is a reduced basic generator matrix.
\\(b) If $C^{\bot H}\subseteq \mathcal{C}$, then the convolutional code $V=\lbrace {\bf u}(D)G(D)|{\bf u}(D)\in \mathbb{F}_{q^{2}}[D]^{n-k}\rbrace$ satisfies $V\subseteq V^{\bot H}$.
\\(c) If $d_{f}$ and $d_{f}^{\bot H}$ denote the free distances of $V$ and $V^{\bot H}$ respectively, $d_{i}$ denotes the minimum distance of the code $\mathcal{C}_{i}=\lbrace {\bf v}\in \mathbb{F}_{q^{2}}^{n}|{\bf v}\tilde{H}_{i}^{t}=0\rbrace$ and $d^{\bot H}$ is the minimum distance of $\mathcal{C}^{\bot H}$, then one has $\min\lbrace d_{0}+d_{\mu}\rbrace\leq d_{f}^{\bot H}\leq d$ and $d_{f}\geq d^{\bot H}$.
\end{theorem}

\subsection{Quantum Convolutional Codes}
Quantum convolutional codes are defined as infinite versions of quantum stabilizer codes. The code is specified by its stabilizer which is a subgroup of the infinite version of the Pauli group that consists of tensor products of generalized Pauli matrices acting on a semi-infinite stream of qudits. The stabilizer can be described by a matrix with polynomial entries
\begin{equation*}
S(D)=(X(D)|Z(D))\in \mathbb{F}_{q}[D]^{(n-k)\times 2n}
\end{equation*}
satisfying $X(D)Z(1/D)^{t}-Z(D)X(1/D)^{t}=0$. A full-rank stabilizer matrix $S(D)$ given above defines a quantum convolutional code $\mathcal{C}$ with parameters $[(n,k,\mu;\gamma,d')]_{q}$, where $n$ is called the frame size, $k$ is the number of logical qudits per frame and $k/n$ is the rate of $\mathcal{C}$. It can be used to encode a sequence of blocks with $k$ qudits in each block into a sequence of blocks with $n$ qudits. The memory of the code is defined as $\mu =\max_{1\leq i\leq n-k,1\leq j\leq n}\lbrace \max \lbrace \textup{deg}~X_{ij}(D),\textup{deg}~Z_{ij}(D)\rbrace\rbrace$. And $d'$ denotes the free distance, $\gamma$ denotes the degree.

In the sequel, we need the following result about how to construct quantum convolutional stabilizer codes by using classical convolutional codes.
\begin{lemma}{\rm \cite{Aly2007Quantum2}}\label{qcc}
Let $V$ be an $(n,(n-k)/2,\gamma;\mu)_{q^{2}}$ convolutional code satisfying $V\subseteq V^{\bot H}$. Then there exists an $[(n,k,\mu;\gamma,d')]_{q}$ convolutional stabilizer code whose free distance is given by $d'=wt(V^{\bot H}\backslash V)$, which is said to be pure if $d' = wt(V^{\bot H})$.
\end{lemma}
\begin{lemma}{\rm \cite{Aly2007Quantum,Guardia2013On}}\label{bound}
The free distance of an $[(n,k,\mu;\gamma,d')]_{q}$, $\mathbb{F}_{q^{2}}$-linear pure convolutional stabilizer code is bounded by
\begin{equation*}
d'\leq \frac{n-k}{2}(\lfloor\frac{2\gamma}{n+k}\rfloor+1)+\gamma+1.
\end{equation*}
\end{lemma}
A quantum convolutional code achieving this bound is called a quantum MDS convolutional code.

\subsection{Generalized Reed-Solomon codes}
Now we recall the basics of GRS codes. Let $n$ be any integer with $1 \leq n\leq q$. Choose ${\bf a}=(a_{0},\cdots,a_{n-1})$ to be an n-tuple of distinct elements of $\mathbb{F}_{q}$, and ${\bf v}=(v_{0},\cdots,v_{n-1})$ to be an $n$-tuple of nonzero elements of $\mathbb{F}_{q}$. Let $k$ be an integer with $1\leq k \leq n$. Then the codes
\begin{equation*}
GRS_{k}({\bf a},{\bf v})=\lbrace(v_{0}f(a_{0}),v_{1}f(a_{1}),\cdots,v_{n-1}f(a_{n-1})) | f\in \mathcal{P}_{k} \rbrace,
\end{equation*}
where $\mathcal{P}_{k}$ denote the set of polynomials of degree less than $k$ in $\mathbb{F}_{q}[x]$, are the GRS codes. It is well known that a GRS code $GRS_{k}({\bf a},{\bf v})$ is an MDS code with parameters $[n,k,n-k+1]_{q}$.

A generator matrix of $GRS_{k}({\bf a},{\bf v})$ is
\begin{equation*}
G=
\left(
  \begin{array}{cccc}
    v_{0} & v_{1} & \cdots & v_{n-1} \\
    v_{0}a_{0} & v_{1}a_{1} & \cdots & v_{n-1}a_{n-1} \\
    v_{0}a_{0}^{2} & v_{1}a_{1}^{2} & \cdots & v_{n-1}a_{n-1}^{2} \\
    \vdots & \vdots & \vdots & \vdots \\
    v_{0}a_{0}^{k-1} & v_{1}a_{1}^{k-1} & \cdots & v_{n-1}a_{n-1}^{k-1} \\
  \end{array}
\right).
\end{equation*}
And a parity check matrix of $GRS_{k}({\bf a},{\bf v})$ is the generator matrix of $GRS_{n-k}({\bf a},{\bf w})$, where ${\bf w}$ is any nonzero codeword in the 1-dimensional code $GRS_{n-1}({\bf a},{\bf v})^{\bot E}$ and satisfies
\begin{equation*}
\sum_{i=0}^{n-1}w_{i}v_{i}h(a_{i})=0
\end{equation*}
for any polynomial $h \in \mathcal{P}_{n-1}$. Therefore a  parity check matrix for $GRS_{k}({\bf a},{\bf v})$ is
\begin{equation*}\label{pcm}
H=
\left(
  \begin{array}{cccc}
    w_{0} & w_{1} & \cdots & w_{n-1} \\
    w_{0}a_{0} & w_{1}a_{1} & \cdots & w_{n-1}a_{n-1} \\
    w_{0}a_{0}^{2} & w_{1}a_{1}^{2} & \cdots & w_{n-1}a_{n-1}^{2} \\
    \vdots & \vdots & \vdots & \vdots \\
    w_{0}a_{0}^{n-k-1} & w_{1}a_{1}^{n-k-1} & \cdots & w_{n-1}a_{n-1}^{n-k-1} \\
  \end{array}
\right).
\end{equation*}

In order to construct good quantum convolutional codes, we need the following two theorems which collect some known infinite families of Hermitian dual-containing GRS codes.
\begin{theorem}{\rm \cite{ZG15}}\label{thmre}
\begin{enumerate}
  \item[(1)] Let $q$ be an odd prime power with the form $2am+1$. Then for each $1\leq b\leq 2a$, there exists a  $[\frac{b(q^{2}-1)}{2a},\frac{b(q^{2}-1)}{2a}-s,s+1]_{q^{2}}$ Hermitian dual-containing GRS code, where $1\leq s\leq(a+1)m$.
  \item[(2)] Let $q$ be an odd prime power with the form $2am+1$. Then for integers $b,c$ such that $b,c\geq0$ and $1\leq b+c\leq 2a$, there exists a  $[\frac{b(q^{2}-1)}{2a}+c(\frac{q^{2}-1}{2a}-q-1),\frac{b(q^{2}-1)}{2a}+c(\frac{q^{2}-1}{2a}-q-1)-s,s+1]_{q^{2}}$ Hermitian dual-containing GRS code, where $1\leq s\leq(a+1)m-1$.
  \item[(3)] Let $q$ be an odd prime power with the form $2am-1$. Then for each $1\leq b\leq 2a$, there exists a $[\frac{b(q^{2}-1)}{2a},\frac{b(q^{2}-1)}{2a}-s,s+1]_{q^{2}}$ Hermitian dual-containing GRS code, where $1\leq s\leq(a+1)m-2$.
  \item[(4)] Let $q$ be an odd prime power with the form $2am-1$. Then for integers $b,c$ such that $b,c\geq0$ and $1\leq b+c\leq 2a$, there exists a  $[\frac{b(q^{2}-1)}{2a}+c(\frac{q^{2}-1}{2a}-q+1),\frac{b(q^{2}-1)}{2a}+c(\frac{q^{2}-1}{2a}-q+1)-s,s+1]_{q^{2}}$ Hermitian dual-containing GRS code, where $1\leq s\leq(a+1)m-3$.
  \item[(5)] Let $q$ be an odd prime power with the form $2am-1$ where $a$ is an odd integer. Then for integers $c_{1},c_{2},c_{3}$ such that $c_{1},c_{2},c_{3}\geq0,\ 0\leq c_{1}+c_{2}\leq a,\ 0\leq c_{1}+c_{3}\leq a$ and $c_{1}+c_{2}+c_{3}\geq1$, there exists a  $[\frac{(c_{2}+c_{3})(q^{2}-1)}{2a}+c_{1}(\frac{q^{2}-1}{a}-q+1),\frac{(c_{2}+c_{3})(q^{2}-1)}{2a}+c_{1}(\frac{q^{2}-1}{a}-q+1)-s,s+1]_{q^{2}}$ Hermitian dual-containing GRS code, where $1\leq s\leq(a+1)m-2$.
  \item[(6)]  Let $q$ be an odd prime power with the form $q=2ab-1$, where $\textup{gcd}(a,b)=1$ and $a,b$ are odd. Then for integer $c$ such that $1\leq c\leq 2(a+b-1)$, there exists a  $[c(q-1),c(q-1)-s,s+1]_{q^{2}}$ Hermitian dual-containing GRS code, where $1\leq s\leq ab+c_{1}-2$, \[c_{1}=\begin{cases}c;&\textup{ if }1\leq c\leq a+b-1,\\
\lfloor\frac{c}{2}\rfloor;&\textup{ if }a+b\leq c\leq 2(a+b-1).\end{cases}\]
  \item[(7)] Let $q$ be an odd prime power with the form $q=2ab+1$, where $\textup{gcd}(a,b)=1$ and $a,b$ are odd. Then for integer $c$ such that $1\leq c\leq 2(a+b-1)$, there exists a  $[c(q+1),c(q+1)-s,s+1]_{q^{2}}$ Hermitian dual-containing GRS code, where $1\leq s\leq ab+c_{1}$, \[c_{1}=\begin{cases}c;&\textup{ if }1\leq c\leq a+b-1,\\
\lfloor\frac{c}{2}\rfloor;&\textup{ if }a+b\leq c\leq 2(a+b-1).\end{cases}\]
\end{enumerate}
\end{theorem}

\begin{theorem}{\rm \cite{JLLX}}\label{thmre3}
\begin{enumerate}
  \item [(1)] Let $t$ be a divisor of $q^{2}-1$. Then, for any $r\leq (q^{2}-1)/t$ and $s\leq (t-1)/(q+1)$, there exists an $[n,n-s,s+1]_{q^{2}}$ Hermitian dual-containing GRS code for both $n=rt$ and $rt+1$.
  \item [(2)] For any $2\leq n \leq q^{2}$, we write $n=n_{1}+\cdots+n_{t}$ with $1\leq t\leq q$ and $2\leq n_{i}\leq q$ for all $i$. Let $1\leq s\leq \min\lbrace n_{1},\cdots,n_{t}\rbrace/2$. Then there exists an $[n,n-s,s+1]_{q^{2}}$ Hermitian dual-containing GRS code.
\end{enumerate}
\end{theorem}

\section{Constructions}\label{seccon}
It is well known that a suitable submatrix of the parity check matrix of a GRS code still corresponds to a GRS code and there are many available choices for such submatrices. Because of this nice property, we are able to construct quantum MDS convolutional codes from GRS codes. We propose two new constructions of quantum MDS convolutional codes in this section.

\subsection{Quantum MDS convolutional codes with $\mu=1$}
\begin{theorem}\label{thm1}
Let $\mathcal{C}$ be a Hermitian dual-containing $[n,k,d]_{q^{2}}$ GRS code, $k\neq n/2$. Then there exist quantum MDS convolutional codes with parameters $[(n,n-2t_{0},1;n-k-t_{0},n-k+1)]_{q}$, where $(n-k)/2\leq t_{0}<n-k$.
\begin{proof}
Suppose $H$ is the parity check matrix of $\mathcal{C}$ and split it into two disjoint submatrices such that
\begin{equation*}
H=
\left(
  \begin{array}{c}
    H_{0}   \\
    H_{1} \\
  \end{array}
\right),
\end{equation*}
where $H_{0}$ has $t_{0}$ rows and $H_{1}$ has $t_{1}=n-k-t_{0}$ rows. Since $(n-k)/2\leq t_{0}<n-k$, we have $t_{0}\geq t_{1}$. It is obvious that $H_{i}$ is still a parity check matrix of an $[n,n-t_{i},t_{i}+1]_{q^{2}}$ GRS code, for $i=0,1$. According to Theorem \ref{gen}, we obtain a convolutional code $V$ that is generated by the reduced basic generator matrix
\begin{equation*}
G(D)=\tilde{H}_{0}+\tilde{H}_{1}D,
\end{equation*}
where $\tilde{H}_{0}=H_{0}$ and $\tilde{H}_{1}$ is derived from $H_{1}$ by adding zero-rows at the bottom such that the number of rows of $\tilde{H}_{1}$ is exactly equal to the number of rows of $\tilde{H}_{0}$. It follows from Theorem \ref{gen} that $V$ is a convolutional code of dimension $t_{0}$, degree $n-k-t_{0}$, memory $1$ and free distance $\geq k+1$. For the free distance of $V^{\bot H}$, we have $\min\lbrace t_{0}+t_{1}+2,n-k+1\rbrace\leq d_{f}^{\bot H}\leq n-k+1$ which forces to $d_{f}^{\bot H}=n-k+1$.

Besides, we have $\mathcal{C}^{\bot H}\subseteq \mathcal{C}$ which gives $V\subseteq V^{\bot H}$ by Lemma \ref{qcc}. Thus we obtain an $[(n,n-2t_{0},1;n-k-t_{0},d')]_{q}$ quantum convolutional code $W$, where $d'=wt(V^{\bot H}\backslash V)$. From Lemma \ref{bound}, we have
\begin{eqnarray*}
d'&\leq&\frac{n-k}{2}(\lfloor\frac{2\gamma}{n+k}\rfloor+1)+\gamma+1\\
  &\leq&t_{0}(\lfloor\frac{n-k-t_{0}}{n-t_{0}}\rfloor+1)+n-k-t_{0}+1\\
  &\leq&n-k+1.
\end{eqnarray*}

Because of the Hermitian dual-containing property of $\mathcal{C}$ and $k\neq n/2$, we have $d_{f}^{\bot H}< d_{f}$. Thus, $d'$ achieves the bound and $W$ is a quantum MDS convolutional code.
\end{proof}
\end{theorem}

It is easy to see that the theorem above is very powerful and can propose about $(n-k)/2$ quantum MDS convolutional codes from each $[n,k,d]_{q^{2}}$ Hermitian dual-containing GRS code. Combining Theorems \ref{thmre}, \ref{thmre3} and \ref{thm1}, we have the following nine new families of quantum MDS convolutional codes.
\begin{theorem}\label{thmre1}
\begin{enumerate}
  \item[(1)] Let $q$ be an odd prime power with the form $2am+1$. Then for each $1\leq b\leq 2a$, there exists a  $[(\frac{b(q^{2}-1)}{2a},\frac{b(q^{2}-1)}{2a}-2t_{0},1;s-t_{0},s+1)]_{q}$ quantum MDS convolutional code, where $1\leq s\leq(a+1)m$, $s/2\leq t_{0}<s$.
  \item[(2)] Let $q$ be an odd prime power with the form $2am+1$. Then for integers $b,c$ such that $b,c\geq0$ and $1\leq b+c\leq 2a$, there exists a  $[(\frac{b(q^{2}-1)}{2a}+c(\frac{q^{2}-1}{2a}-q-1),\frac{b(q^{2}-1)}{2a}+c(\frac{q^{2}-1}{2a}-q-1)-2t_{0},1;s-t_{0},s+1)]_{q}$ quantum MDS convolutional code, where $1\leq s\leq(a+1)m-1$, $s/2\leq t_{0}<s$.
  \item[(3)] Let $q$ be an odd prime power with the form $2am-1$. Then for each $1\leq b\leq 2a$, there exists a $[(\frac{b(q^{2}-1)}{2a},\frac{b(q^{2}-1)}{2a}-2t_{0},1;s-t_{0},s+1)]_{q}$ quantum MDS convolutional code, where $1\leq s\leq(a+1)m-2$, $s/2\leq t_{0}<s$.
  \item[(4)] Let $q$ be an odd prime power with the form $2am-1$. Then for integers $b,c$ such that $b,c\geq0$ and $1\leq b+c\leq 2a$, there exists a  $[(\frac{b(q^{2}-1)}{2a}+c(\frac{q^{2}-1}{2a}-q+1),\frac{b(q^{2}-1)}{2a}+c(\frac{q^{2}-1}{2a}-q+1)-2t_{0},1;s-t_{0},s+1)]_{q}$ quantum MDS convolutional code, where $1\leq s\leq(a+1)m-3$, $s/2\leq t_{0}<s$.
  \item[(5)] Let $q$ be an odd prime power with the form $2am-1$ where $a$ is an odd integer. Then for integers $c_{1},c_{2},c_{3}$ such that $c_{1},c_{2},c_{3}\geq0,\ 0\leq c_{1}+c_{2}\leq a,\ 0\leq c_{1}+c_{3}\leq a$ and $c_{1}+c_{2}+c_{3}\geq1$, there exists a  $[(\frac{(c_{2}+c_{3})(q^{2}-1)}{2a}+c_{1}(\frac{q^{2}-1}{a}-q+1),\frac{(c_{2}+c_{3})(q^{2}-1)}{2a}+c_{1}(\frac{q^{2}-1}{a}-q+1)-2t_{0},1;s-t_{0},s+1)]_{q}$ quantum MDS convolutional code, where $1\leq s\leq(a+1)m-2$, $s/2\leq t_{0}<s$.
  \item[(6)]  Let $q$ be an odd prime power with the form $q=2ab-1$, where $\textup{gcd}(a,b)=1$ and $a,b$ are odd. Then for integer $c$ such that $1\leq c\leq 2(a+b-1)$, there exists a  $[(c(q-1),c(q-1)-2t_{0},1;s-t_{0},s+1)]_{q}$ quantum MDS convolutional code, where $1\leq s\leq ab+c_{1}-2$, $s/2\leq t_{0}<s$, \[c_{1}=\begin{cases}c;&\textup{ if }1\leq c\leq a+b-1,\\
\lfloor\frac{c}{2}\rfloor;&\textup{ if }a+b\leq c\leq 2(a+b-1).\end{cases}\]
  \item[(7)] Let $q$ be an odd prime power with the form $q=2ab+1$, where $\textup{gcd}(a,b)=1$ and $a,b$ are odd. Then for integer $c$ such that $1\leq c\leq 2(a+b-1)$, there exists a  $[(c(q+1),c(q+1)-2t_{0},1;s-t_{0},s+1)]_{q}$ quantum MDS convolutional code, where $1\leq s\leq ab+c_{1}$, $s/2\leq t_{0}<s$, \[c_{1}=\begin{cases}c;&\textup{ if }1\leq c\leq a+b-1,\\
\lfloor\frac{c}{2}\rfloor;&\textup{ if }a+b\leq c\leq 2(a+b-1).\end{cases}\]
  \item [(8)] Let $t$ be a divisor of $q^{2}-1$. Then, for any $r\leq (q^{2}-1)/t$ and $s\leq (t-1)/(q+1)$, there exists an $[(n,n-2t_{0},1;s-t_{0},s+1)]_{q}$ quantum MDS convolutional code for both $n=rt$ and $rt+1$, where $s/2\leq t_{0}<s$.
  \item [(9)] For any $2\leq n \leq q^{2}$, we write $n=n_{1}+\cdots+n_{t}$ with $1\leq t\leq q$ and $2\leq n_{i}\leq q$ for all $i$. Let $1\leq s\leq \min\lbrace n_{1},\cdots,n_{t}\rbrace/2$. Then there exists an $[(n,n-2t_{0},1;s-t_{0},s+1)]_{q}$ quantum MDS convolutional code, where $s/2\leq t_{0}<s$.
\end{enumerate}
\end{theorem}

Table \ref{table1} lists some quantum MDS convolutional codes obtained from Theorem \ref{thmre1}.

\begin{table}
\begin{center}
\caption{Quantum MDS Convolutional Codes}\label{table1}
\begin{tabular}{|c|c|c|c|c|c|c|}
  \hline
  $q$  & $a$ & $b$ & $c$         & $[(n,k,\mu;\gamma,d')]_{q}$             & $s$              & $t_{0}$ \\
  \hline
  $17$ & $1$ & $2$ & $\diagdown$ & $[(288,288-2t_{0},1;s-t_{0},s+1)]_{17}$ & $1\leq s\leq 16$ & $s/2\leq t_{0}<s$ \\
  \hline
  $17$ & $2$ & $2$ & $1$ & $[(198,198-2t_{0},1;s-t_{0},s+1)]_{17}$ & $1\leq s\leq 11$ & $s/2\leq t_{0}<s$ \\
  \hline
  $11$ & $2$ & $4$ & $\diagdown$ & $[(120,120-2t_{0},1;s-t_{0},s+1)]_{11}$ & $1\leq s\leq 7$ & $s/2\leq t_{0}<s$ \\
  \hline
  $23$ & $3$ & $3$ & $2$ & $[(394,394-2t_{0},1;s-t_{0},s+1)]_{23}$ & $1\leq s\leq 13$ & $s/2\leq t_{0}<s$ \\
  \hline
  $29$ & $3$ & $5$ & $10$ & $[(300,300-2t_{0},1;s-t_{0},s+1)]_{29}$ & $1\leq s\leq 20$ & $s/2\leq t_{0}<s$ \\
  \hline
  $31$ & $3$ & $5$ & $10$ & $[(320,320-2t_{0},1;s-t_{0},s+1)]_{31}$ & $1\leq s\leq 20$ & $s/2\leq t_{0}<s$ \\
  \hline
\end{tabular}
\end{center}
\end{table}
\newpage
\subsection{Quantum MDS convolutional codes with $\mu=2$}
The following is a similar construction with memory and degree both equal two.
\begin{theorem}\label{thm2}
Let $\mathcal{C}$ be a Hermitian dual-containing $[n,k,d]_{q^{2}}$ GRS code, $n/2<k<n-2$. Then there exist a quantum MDS convolutional code with parameters $[(n,2k-n+4,2;2,n-k+1)]_{q}$.
\begin{proof}
Suppose $H$ is the parity check matrix of $\mathcal{C}$ and split it into three disjoint submatrices such that
\begin{equation*}
H=
\left(
  \begin{array}{c}
    H_{0}   \\
    H_{1} \\
    H_{2} \\
  \end{array}
\right),
\end{equation*}
where $H_{0}$ has $t_{0}=n-k-2$ rows and both $H_{1}$ and $H_{2}$ have only one row. It is obvious to see that $H_{0}$ is the parity check matrix of an $[n,k+2,n-k-1]_{q^{2}}$ GRS code and $H_{i}$ is the parity check matrix of an $[n,n-1,2]_{q^{2}}$ GRS code, for $i=1,2$.  According to Theorem \ref{gen}, we obtain a convolutional code $V$ that is generated by the reduced basic generator matrix
\begin{equation*}
G(D)=\tilde{H}_{0}+\tilde{H}_{1}D+\tilde{H}_{2}D^{2},
\end{equation*}
where $\tilde{H}_{0}=H_{0}$ and $\tilde{H}_{1}$ and $\tilde{H}_{2}$ are derived from $H_{1}$ and $H_{2}$ respectively, by adding zero-rows at the bottom such that the numbers of rows of $\tilde{H}_{1}$ and $\tilde{H}_{2}$ are exactly equal to the number of rows of $\tilde{H}_{0}$. It follows from Theorem \ref{gen} that $V$ is a convolutional code of dimension $t_{0}$, degree $2$, memory $2$ and free distance $\geq k+1$. For the free distance of $V^{\bot H}$, we have $\min\lbrace t_{0}+t_{2}+2,n-k+1\rbrace\leq d_{f}^{\bot H}\leq n-k+1$ which forces to $d_{f}^{\bot H}=n-k+1$.

Besides, we have $\mathcal{C}^{\bot H}\subseteq \mathcal{C}$ which gives $V\subseteq V^{\bot H}$ by Lemma \ref{qcc}. Thus we obtain an $[(n,2k-n+4,2;2,d')]_{q}$ quantum convolutional code $W$, where $d'=wt(V^{\bot H}\backslash V)$. It follows from Lemma \ref{bound} that
\begin{eqnarray*}
d'&\leq&\frac{n-k}{2}(\lfloor\frac{2\gamma}{n+k}\rfloor+1)+\gamma+1\\
  &\leq&(n-k-2)(\lfloor\frac{4}{2k+4}\rfloor+1)+3\\
  &\leq&n-k+1.
\end{eqnarray*}

Because of the Hermitian dual-containing property of $\mathcal{C}$ and $k\neq n/2$, we have $d_{f}^{\bot H}< d_{f}$. Thus, $d'$ achieves the bound and $W$ is a quantum MDS convolutional code.
\end{proof}
\end{theorem}

By Theorems \ref{thmre}, \ref{thmre3} and \ref{thm2}, we have the following nine new families of quantum MDS convolutional codes.
\begin{theorem}\label{thmre2}
\begin{enumerate}
  \item[(1)] Let $q$ be an odd prime power with the form $2am+1$. Then for each $1\leq b\leq 2a$, there exists a  $[(\frac{b(q^{2}-1)}{2a},\frac{b(q^{2}-1)}{2a}-2s+4,2;2,s+1)]_{q}$ quantum MDS convolutional code, where $3\leq s\leq(a+1)m$.
  \item[(2)] Let $q$ be an odd prime power with the form $2am+1$. Then for integers $b,c$ such that $b,c\geq0$ and $1\leq b+c\leq 2a$, there exists a  $[(\frac{b(q^{2}-1)}{2a}+c(\frac{q^{2}-1}{2a}-q-1),\frac{b(q^{2}-1)}{2a}+c(\frac{q^{2}-1}{2a}-q-1)-2s+4,2;2,s+1)]_{q}$ quantum MDS convolutional code, where $3\leq s\leq(a+1)m-1$.
  \item[(3)] Let $q$ be an odd prime power with the form $2am-1$. Then for each $1\leq b\leq 2a$, there exists a $[(\frac{b(q^{2}-1)}{2a},\frac{b(q^{2}-1)}{2a}-2s+4,2;2,s+1)]_{q}$ quantum MDS convolutional code, where $3\leq s\leq(a+1)m-2$.
  \item[(4)] Let $q$ be an odd prime power with the form $2am-1$. Then for integers $b,c$ such that $b,c\geq0$ and $1\leq b+c\leq 2a$, there exists a  $[(\frac{b(q^{2}-1)}{2a}+c(\frac{q^{2}-1}{2a}-q+1),\frac{b(q^{2}-1)}{2a}+c(\frac{q^{2}-1}{2a}-q+1)-2s+4,2;2,s+1)]_{q}$ quantum MDS convolutional code, where $3\leq s\leq(a+1)m-3$.
  \item[(5)] Let $q$ be an odd prime power with the form $2am-1$ where $a$ is an odd integer. Then for integers $c_{1},c_{2},c_{3}$ such that $c_{1},c_{2},c_{3}\geq0,\ 0\leq c_{1}+c_{2}\leq a,\ 0\leq c_{1}+c_{3}\leq a$ and $c_{1}+c_{2}+c_{3}\geq1$, there exists a  $[(\frac{(c_{2}+c_{3})(q^{2}-1)}{2a}+c_{1}(\frac{q^{2}-1}{a}-q+1),\frac{(c_{2}+c_{3})(q^{2}-1)}{2a}+c_{1}(\frac{q^{2}-1}{a}-q+1)-2s+4,2;2,s+1)]_{q}$ quantum MDS convolutional code, where $3\leq s\leq(a+1)m-2$.
  \item[(6)]  Let $q$ be an odd prime power with the form $q=2ab-1$, where $\textup{gcd}(a,b)=1$ and $a,b$ are odd. Then for integer $c$ such that $1\leq c\leq 2(a+b-1)$, there exists a  $[(c(q-1),c(q-1)-2s+4,2;2,s+1)]_{q}$ quantum MDS convolutional code, where $3\leq s\leq ab+c_{1}-2$, \[c_{1}=\begin{cases}c;&\textup{ if }1\leq c\leq a+b-1,\\
\lfloor\frac{c}{2}\rfloor;&\textup{ if }a+b\leq c\leq 2(a+b-1).\end{cases}\]
  \item[(7)] Let $q$ be an odd prime power with the form $q=2ab+1$, where $\textup{gcd}(a,b)=1$ and $a,b$ are odd. Then for integer $c$ such that $1\leq c\leq 2(a+b-1)$, there exists a  $[(c(q+1),c(q+1)-2s+4,2;2,s+1)]_{q}$ quantum MDS convolutional code, where $3\leq s\leq ab+c_{1}$, \[c_{1}=\begin{cases}c;&\textup{ if }1\leq c\leq a+b-1,\\
\lfloor\frac{c}{2}\rfloor;&\textup{ if }a+b\leq c\leq 2(a+b-1).\end{cases}\]
  \item [(8)] Let $t$ be a divisor of $q^{2}-1$. Then, for any $r\leq (q^{2}-1)/t$ and $s\leq (t-1)/(q+1)$, there exists an $[(n,n-2s+4,2;2,s+1)]_{q}$ quantum MDS convolutional code for both $n=rt$ and $rt+1$.
  \item [(9)] For any $2\leq n \leq q^{2}$, we write $n=n_{1}+\cdots+n_{t}$ with $1\leq t\leq q$ and $2\leq n_{i}\leq q$ for all $i$. Let $1\leq s\leq \min\lbrace n_{1},\cdots,n_{t}\rbrace/2$. Then there exists an $[(n,n-2s+4,2;2,s+1)]_{q}$ quantum MDS convolutional code.
\end{enumerate}
\end{theorem}

Table \ref{table2} lists some quantum MDS convolutional codes obtained from Theorem \ref{thmre2}.

\begin{table}
\begin{center}
\caption{Quantum MDS Convolutional Codes}\label{table2}
\begin{tabular}{|c|c|c|c|c|c|}
  \hline
  $q$  & $a$ & $b$ & $c$         & $[(n,k,\mu;\gamma,d')]_{q}$             & $s$              \\
  \hline
  $17$ & $1$ & $2$ & $\diagdown$ & $[(288,292-2s,2;2,s+1)]_{17}$ & $3\leq s\leq 16$ \\
  \hline
  $17$ & $2$ & $2$ & $1$ & $[(198,202-2s,2;2,s+1)]_{17}$ & $3\leq s\leq 11$  \\
  \hline
  $11$ & $2$ & $4$ & $\diagdown$ & $[(120,124-2s,2;2,s+1)]_{11}$ & $3\leq s\leq 7$ \\
  \hline
  $23$ & $3$ & $3$ & $2$ & $[(394,398-2s,2;2,s+1)]_{23}$ & $3\leq s\leq 13$ \\
  \hline
  $29$ & $3$ & $5$ & $10$ & $[(300,304-2s,2;2,s+1)]_{29}$ & $3\leq s\leq 20$ \\
  \hline
  $31$ & $3$ & $5$ & $10$ & $[(320,324-2s,2;2,s+1)]_{31}$ & $3\leq s\leq 20$  \\
  \hline
\end{tabular}
\end{center}
\end{table}

\section{Conclusion}\label{conclusion}
 In this paper, we show that we can get quantum MDS convolutional codes from any GRS codes that are Hermitian dual-containing. In particular, we present two new constructions of quantum MDS convolutional codes and propose eighteen new classes of quantum MDS convolutional codes, providing a wide range of parameters.


\begin{thebibliography}{10}

\bibitem{Aly2007Quantum2}
S.~A. Aly, M.~Grassl, A.~Klappenecker, M.~Rotteler, and P.~K. Sarvepalli.
\newblock Quantum convolutional {BCH} codes.
\newblock In {\em Int. Symp. Inform. Theory, ISIT}, pages 180--183, 2007.

\bibitem{Aly2007Quantum}
S.~A. Aly, A.~Klappenecker, and P.~K. Sarvepalli.
\newblock Quantum convolutional codes derived from generalized {R}eed-{S}olomon
  codes.
\newblock In {\em Int. Symp. Inform. Theory, ISIT}, pages 821--825, 2007.

\bibitem{def1}
H.~F. Chau.
\newblock Quantum convolutional error-correcting codes.
\newblock {\em Phys. Rev. A {\rm (3)}}, 58(2):905--909, 1998.

\bibitem{CLHL}
J.~Chen, J.~Li, Y.~Huang, and J.~Lin.
\newblock Some families of asymmetric quantum codes and quantum convolutional
  codes from constacyclic codes.
\newblock {\em Linear Algebra Appl.}, 475:186--199, 2015.

\bibitem{De2004A}
A.~C.~A. De~Almeida and R.~Palazzo~Jr.
\newblock A concatenated [(4, 1, 3)] quantum convolutional code.
\newblock In {\em IEEE Inform.Theory Workshop (ITW)}, pages 28--33, 2004.

\bibitem{EJKL}
M.~F. Ezerman, S.~Jitman, H.~M. Kiah, and S.~Ling.
\newblock Pure asymmetric quantum {MDS} codes from {CSS} construction: a
  complete characterization.
\newblock {\em Int. J. Quantum Inf.}, 11(3):1350027, 10, 2013.

\bibitem{Forney2007Convolutional}
G.~D. Forney, M.~Grassl, and S.~Guha.
\newblock Convolutional and tail-biting quantum error-correcting codes.
\newblock {\em IEEE Trans. Inform. Theory}, 53(3):865--880, 2007.

\bibitem{Grassl2006Non}
M.~Grassl and M.~Rotteler.
\newblock Non-catastrophic encoders and encoder inverses for quantum
  convolutional codes.
\newblock In {\em Int. Symp. Inform. Theory, ISIT}, pages 1109--1113, 2006.

\bibitem{Grassl2007Constructions}
M.~Grassl and M.~Rotteler.
\newblock Constructions of quantum convolutional codes.
\newblock In {\em Int. Symp. Inform. Theory, ISIT}, pages 816--820, 2007.

\bibitem{Grassl2007Quantum}
M.~Grassl and M.~Rotteler.
\newblock Quantum block and convolutional codes from self-orthogonal product
  codes.
\newblock {\em Int. Symp. Inform. Theory, ISIT}, pages 1018--1022, 2007.

\bibitem{FUN}
W.~C. Huffman and V.~Pless.
\newblock {\em Fundamentals of error-correcting codes}.
\newblock Cambridge University Press, Cambridge, 2003.

\bibitem{JLLX}
L.~Jin, S.~Ling, J.~Luo, and C.~Xing.
\newblock Application of classical {H}ermitian self-orthogonal {MDS} codes to
  quantum {MDS} codes.
\newblock {\em IEEE Trans. Inform. Theory}, 56(9):4735--4740, 2010.

\bibitem{Giuliano2014On}
G.~G. La~Guardia.
\newblock On classical and quantum {MDS}-convolutional {BCH} codes.
\newblock {\em IEEE Trans. Inform. Theory}, 60(1):304--312, 2014.

\bibitem{Guardia2013On}
G.~G. La~Guardia.
\newblock On negacyclic {MDS}-convolutional codes.
\newblock {\em Linear Algebra Appl.}, 448:85--96, 2014.

\bibitem{Lidar2013Quantum}
D.~A. Lidar and T.~A. Brun.
\newblock {\em Quantum error correction}.
\newblock Cambridge University Press, 2013.

\bibitem{Harold2003Description}
H.~Ollivier and J.~P. Tillich.
\newblock Description of a quantum convolutional code.
\newblock {\em Phys. Rev. Lett.}, 91(17):177902, 2003.

\bibitem{Ollivier2004Quantum}
H.~Ollivier and J.~P. Tillich.
\newblock Quantum convolutional codes: fundamentals.
\newblock {\em HAL-INRIA}, 54(9):4053--4068, 2004.

\bibitem{QJZ}
J.~Qian and L.~Zhang.
\newblock Constructions of new quantum burst-correcting codes.
\newblock {\em Internat. J. Theoret. Phys.}, 54(3):917--926, 2015.

\bibitem{Tan2010Efficient}
P.~Tan and J.~Li.
\newblock Efficient quantum stabilizer codes: {LDPC} and {LDPC}-convolutional
  constructions.
\newblock {\em IEEE Trans. Inform. Theory}, 56(1):476--491, 2010.

\bibitem{Wilde2007Entanglement}
M.~M. Wilde and T.~A. Brun.
\newblock Entanglement-assisted quantum convolutional coding.
\newblock {\em Phys. Rev. A {\rm (3)}}, 81(4):042333, 21, 2010.

\bibitem{Wilde2010Convolutional}
M.~M. Wilde, H.~Krovi, and T.~A. Brun.
\newblock Convolutional entanglement distillation.
\newblock In {\em Int. Symp. Inform. Theory, ISIT}, pages 2657--2661, 2010.

\bibitem{ZCL}
G.~Zhang, B.~Chen, and L.~Li.
\newblock A construction of {MDS} quantum convolutional codes.
\newblock {\em Internat. J. Theoret. Phys.}, 54(9):3182--3194, 2015.

\bibitem{ZG15}
T.~Zhang and G.~Ge.
\newblock Quantum {MDS} codes with large minimum distance.
\newblock submitted.

\end{thebibliography}
\end{document}